# Experimental observation of anisotropic Adler-Bell-Jackiw anomaly in type-II Weyl semimetal WTe$_{1.98}$ crystals at the quasi-classical regime

Yang-Yang Lv,[1,†] Xiao Li,[2,†] Bin-Bin Zhang,[1,†] W. Y. Deng,[2] Shu-Hua Yao,[1,*] Y. B. Chen,[2,*] Jian Zhou,[1,*] Shan-Tao Zhang,[1] Ming-Hui Lu,[1] Lei Zhang,[4] Mingliang Tian,[3,4] L. Sheng[2,3] and Yan-Feng Chen[1,3]

[1]National Laboratory of Solid State Microstructures & Department of Materials Science and Engineering, Nanjing University, Nanjing, Jiangsu 210093 China

[2]National Laboratory of Solid State Microstructures & Department of Physics, Nanjing University, Nanjing, Jiangsu 210093 China

[3]Collaborative Innovation Center of Advanced Microstructure, Nanjing University, Nanjing, Jiangsu 210093 China

[4]High Magnetic Field Laboratory, Chinese Academy of Sciences, Hefei, Anhui 230031 China

†Yang-Yang Lv, Xiao Li and Bin-Bin Zhang contribute equally.

Corresponding authors: S. H. Y, shyao@nju.edu.cn; Y. B. C, ybchen@nju.edu.cn; J. Z, zhoujian@nju.edu.cn

The asymmetric electron dispersion in type-II Weyl semimetal theoretically hosts anisotropic transport properties. Here we observe the significant anisotropic Adler-Bell-Jackiw (ABJ) anomaly in the Fermi-level delicately adjusted WTe$_{1.98}$ crystals. Quantitatively, $C_W$, a coefficient representing intensity of ABJ anomaly, along *a*- and *b*-axis of WTe$_{1.98}$ are 0.030 and 0.051 T$^{-2}$ at 2 K, respectively. We found that temperature-sensitive ABJ anomaly is attributed to topological phase transition from type-II Weyl semimetal to trivial semimetal, which is verified by first-principles calculation using experimentally determined lattice parameters at different temperatures. Theoretical electrical transport study reveals that observation of ansotropic ABJ both along *a*- and *b*-axis in WTe$_{1.98}$ is attributed to electrical transport in the quasi-classical regime. Our work may suggest that electron-doped WTe$_2$ is an ideal playground to explore the novel properties in type-II Weyl semimetals.

Adler-Bell-Jackiw (ABJ) anomaly is a remarkable phenomenon that originates from breaking of chiral symmetry in the massless Weyl fermion under quantum fluctuation [1-6]. Previously, ABJ anomaly is mainly observed in high-energy particle physics [1,2,7], while it was recently observed in Weyl semimetals, for example, TaAs, NbAs, and TaP etc [8-12]. In these materials, negative magnetoresistance (MR) due to ABJ anomaly can be observed at low temperatures when the magnetic field is parallel to the electric field. And this negative MR vanishes when the magnetic field is misaligned with respect to electric field or temperature is above a threshold value.

The latest progress in this field is the extension of the Weyl semimetal to type-II Weyl one in which the effective electron dispersion relationship breaks the honored Lorentz invariance [13-15]. Except the Fermi arc at Fermi surface, a remarkable fingerprint of type-II Weyl semimetals is the anisotropic transport properties, such as anisotropic ABJ anomaly. Currently, several angle-resolved photoemission spectroscopy works have claimed to observe the Fermi arc in type-II Weyl semimetals, such as $WTe_2$ [16-19], $MoTe_2$ [20-25], and LaAlGe [26]. In a recent work, Wang, *et al.*, reported the observation of negative magnetoresistance along *b*-axis in electrical-gated $WTe_2$ *few-layered* samples [27]. A natural question is can we dope electron into $WTe_2$ *crystal* to observe the ABJ anomaly. Obviously, electron-doped $WTe_2$ crystal has permanent ABJ anomaly rather than temperoray one under electrical-gating [27].

In this work, we report the evidence of anisotropic ABJ anomaly in a type-II Weyl semimetal: Fermi-level delicately adjusted $WTe_{1.98}$ crystals. We found that temperature-sensitive ABJ anomaly in $WTe_{1.98}$ crystal is attributed to topological phase transition of type-II Weyl semimetal phase in $WTe_{1.98}$ converting to topologically trivial semimetal at high temperature (~60 K), which is verified by first-principles calculation using experimentally determined lattice parameters at different temperatures. Theoretical electrical transport study reveals that the observation of anisotropic ABJ anomaly along both *a*- and *b*-axis is attributed to the electrical transport at the quasi-classical regime, rather than ultra-quantum regime reported in the work of Wang [27].

We delicately synthesized a series of Te-deficient WTe$_{2-\delta}$ (δ is varied from 0.144 to 0.011) single crystals by the post-anealing described in supplementary information (SI) [28]. Theoretically, the position of Weyl points is about 60 meV above Fermi energy in stoichiometric WTe$_2$ [13,28]. In experiment we can only observe ABJ anomaly in samples with δ around 0.02. In the SI [28], the shift of Fermi energy in WTe$_{1.98}$, determined by comparsion between carrier concentration and integrated electron density of state, is quite approximately close to Weyl points.

As shown in Fig. 1(a), bulk WTe$_2$ has a layered crystal structure. The X-ray diffraction (XRD) pattern of a WTe$_2$ crystal is depicted in Fig. 1(b). Only the reflections of (0 0 2$k$) show up, suggesting that the crystals have $c$-axis orientation. The full-width at half maximum of (002) pole is as small as 0.08°, which infers the high crystalline quality. Fig. 1(c) plots the energy-dispersive spectroscopy (EDS) mappings of W and Te elements, which indicates that the two elements (W and Te) are uniformly distributed in WTe$_2$ crystals. High resolution transmission electron microscopy (TEM) and the corresponding selected area electron diffraction (SAED) pictures (see Fig. 1(d)) prove that the as-grown WTe$_2$ crystals have single crystalline quality at the atomic scale.

Because ABJ anomaly can only be observed in WTe$_{1.98}$, we will focus on describing electrical transport data of WTe$_{1.98}$ in the following paragraph.

The longitudinal resistivity $\rho_{xx}$ of the WTe$_{1.98}$ single crystals, with electrical current along $a$- and $b$-axis, are shown in Fig. 2. Figure 2(a) depicts the temperature-dependent resistivity $\rho_{xx,a}$ under various magnetic fields aligned along $c$-axis, with electrical current along $a$-axis. In the absence of external magnetic field, the resistivity exhibits a typical metallic behavior. But an insulator-like behavior is observed at low temperature when magnetic field is applied. The physical origin of metal-insulator transition in WTe$_2$ is still under hot debate [29]. The resistivity behavior $\rho_{xx,b}$ with electrical current along $b$-axis under magnetic field is quite similar to that along $a$-axis (see Fig. 2(b)).

The angle- and field-dependent MR of WTe$_{1.98}$ crystals under different temperatures, with both electrical current ($\vec{j}$) and magnetic field ($\vec{B}$) applied along either *a*- or *b*-direction, are shown in Fig. 3. One can see from Fig. 3(a) that the negative MR reaches the maximum (-30%) at 2 K and vanishes at about 40 K under $\vec{B}//\vec{j}//a$-axis. Figure 3(b) displays the field-dependence of MR measured at 2 K on different misalignment angle θ (see inset for definition of θ). The MR reaches up to 1200 % when the applied field is perpendicular to the current, but it changes negative when θ is between 0° and 5° (see the inset of Fig. 3(b)). These features are quite similar to those observed in TaAs that is a prototypal type-I Weyl semimetal [8,9]. The similar MR behavior is also observed with $\vec{B}//\vec{j}//b$-axis. As shown in Figs. 3(c) and 3(d), the negative MR of about -40 % is observed at 2 K and vanished at 30 K; and negative MR measured at 2 K is observed when the misalignment angle θ is smaller than 20° (see inset of Fig. 3(d)). Compared Fig. 3(a) to 3(c), the negative MR, measured at 9 T and 2 K, are -30% and -40% in the case of $\vec{B}//\vec{j}//a$ and $\vec{B}//\vec{j}//b$, respectively. It therefore can be concluded that there is an obviously anisotropic negative MR in electron-doped WTe$_2$ crystals. In the SI [28], we rule out the other possibilities, for example current-injecting effect, magnetism and ultra-quantum effect, leading to the observed negative MR effect.

ABJ anomaly induced negative MR in WTe$_{1.98}$ crystals is further quantitatively analyzed by semi-classical formulas that have been successfully employed to analyze the ABJ anomaly in type-I Weyl semimetal [5,9]. The fitted formulas are:

$$\sigma(B) = (1 + C_W B^2)\sigma_{WAL} + \sigma_N \tag{1}$$

$$\sigma_{WAL} = \sigma_o + a\sqrt{B} \tag{2}$$

where $\sigma_{WAL}$ is the conductivity from weak anti-localization corrections associated with spin-orbit coupling and $\sigma_N$ comes from conventional Fermi surface contributions except Weyl points. The most important term in the formula is $C_W B^2$ with a positive constant $C_W$, originating from ABJ anomaly. Firstly, the negative MR along both *a*-axis and *b*-axis measured at 2 K can be well fitted (shown in Fig. 4(a)) with the fitting parameters summarized in table 1. The $C_W$ in *a*-axis and *b*-axis are 0.030 and 0.051 T$^{-2}$ at 2 K, respectively, which obviously indicate that the ABJ anomaly along *b*-axis is more significant than that along *a*-axis. Secondly, the temperature dependent negative MR, with $\vec{B}//\vec{j}//b$, are also fitted in Fig. 4(b). One can see the overall agreement between fitted and experimental results. The fitting parameters, summarized in SI [28], suggest that ABJ anomaly effect is decreased with increased temperature. Quantitatively, $C_W$ is decreased from 0.051 to 0.007 T$^{-2}$ when the temperature is increased from 2 to 20 K. Temperature-sensitive ABJ anomaly is also observed in prototypal type-I Weyl semimetal-TaAs [8,9].

As discussed in Fig. 3, ABJ anomaly of WTe$_{1.98}$ crystals can only be observed below 40 or 30 K dependent on direction. Here we discuss two possible factors leading to the temperature-sensitive ABJ anomaly in WTe$_2$. The first factor is the possible temperature-dependent topological phase transition in WTe$_2$. In other words, type-II Weyl semimetal phase in WTe$_2$ is strongly dependent on lattice constants (equivalent to temperatures). Firstly, we measured the temperature dependent powder XRD of WTe$_2$ at 35, 60, 100, 200, and 300 K. Then their crystal structures at these temperatures were determined by Rietveld refinement. Figure 5(a) displays the agreement between observed (crosses) and fitted (solid lines) diffraction patterns at 35 K. All the refined lattice constants are summarized in Table S2 in SI [28], and they can be well linearly fitted shown in Fig. 5(b). Based on the linear fit, we then calculated the evolution of Weyl points in WTe$_2$ crystal at different lattice constants from 0 to 130 K (details can be found in SI [28]). Figure 5(c) shows two band crossing points (Weyl points) at 0 K, which is very similar to the result of Soluyanov [13]. In other words, WTe$_2$ at 0 K is a type-II Weyl semimetal. However, at a higher temperature of 130 K (see Fig. 5(d)), there is an energy gap (about 1.9 meV) which indicates that WTe$_2$ at 130 K is not a type-II Weyl semimetal any more. The whole evolution from the type-II Weyl semimetal to the trivial semimetal in WTe$_2$ can be clearly seen in Fig. 5(e), in which we find that the separation distance between the two Weyl points decreases to almost zero at about 70 K, while at the same time, the obvious energy gap appears when the temperature exceeds about 70 K. Therefore, we can conclude that the WTe2 undergoes a topological phase transition from the type-II Weyl semimetal to the trivial one with the transition temperature of about 70 K. These calculations clearly deomonstrate the highly sensitive Weyl state in WTe2 on the lattice constants (equivalent to the temperature). Thus, the above-mentioned temperature-sensitive ABJ anomaly should be attributed to topological phase transition in WTe2 under thermal perturbation. The second factor leading to temperature-sensitive ABJ anomaly is the Fermi level changed by the thermal agitation. Based on current data, we cannot distinguish which factor is dominated to the temperature-sensitive ABJ anomaly in WTe2 crystal.

In the following paragraphs, we will answer two important questions. One is how can we correlated the calculated electronic band structure with observed ABJ anomaly in a semi-quantitatively/quantiatively way? The other is how to reconcile our observation of ABJ anomaly on both *a*- and *b*-axis, while it is only observed along *b*-axis in electrical-gated WTe2 few-layered samples as reported [27]? We analyze the anisotropic chiral anomaly, which leads to the anisotropic negative longitudinal magneto-resistance. The general form of the Hamiltonian around a Weyl point is [13]

$$H(\vec{k}) = Ak_x + Bk_y + (ak_x + ck_y)\sigma_y + (bk_x + dk_y)\sigma_z + ek_z\sigma_x. \quad (3)$$

The Berry curvature is calculated to be

$$\vec{\Omega}_{n,k} = -\frac{n}{2g^3}(ad - bc)e\vec{k}, \quad (4)$$

where $g = \sqrt{e^2 k_z^2 + (ak_x + ck_y)^2 + (bk_x + dk_y)^2}$, $n = \pm 1$ for n-th band.

In the ultra-quantum limit, the chiral anomaly is characterized by the ratio $R$ around the Weyl point, defined as [13]

$$R = \frac{(Ak_x + Bk_y)^2}{g^2}. \quad (5)$$

Chiral anomaly and related negative magnetoresistance occur ONLY in the direction with $R > 1$ [13]. It is convenient to express $R$ defined in Eq. (3) along the *a* or *b*

axis as

$$R = \left(\frac{\lambda_+ + \lambda_-}{\lambda_+ - \lambda_-}\right)^2, \quad (6)$$

where $\lambda_\pm$ are the slopes of the two energy dispersion curves at the Weyl point along the $a$ or $b$ axis. It is easy to find that $R > 1$, if the two slopes have the same sign, and $R < 1$ otherwise. By using the data of the slopes around the two Weyl points obtained from our first principle calculation, we find $R > 1$ occurs only along the $b$ axis and $R < 1$ along the $a$ axis, as shown in the Fig. S10 [28]. This means that in the ultra-quantum regime, the chiral anomaly induced negative longitudinal magnetoresistance only happens along the $b$ axis.

However, in the opposite classical limit, it is found that chiral anomaly induced negative magnetoresistance is isotropic [30], and so positive longitudinal magneto-conductivity is present along any arbitrary direction. The longitudinal conductivity $\sigma_{uu}$ can be obtained based on the classical Boltzmann equation, incorporating Berry curvature effect, as [30]

$$\sigma_{uu} = \sigma_{u,0} + \gamma_u B^2, \quad (7)$$

with

$$\sigma_{u,0} = e^2 \tau \sum_n \int dk^3 D_{u,n} v_u^2 \left(-\frac{\partial f_{eq}}{\partial \varepsilon}\right), \quad (8)$$

$$\gamma_u = \frac{e^4 \tau}{\hbar^2} \sum_n \int dk^3 D_{u,n} \left(\vec{\Omega}_{n,k} \cdot \vec{v}_{n,k}\right)^2 \left(-\frac{\partial f_{eq}}{\partial \varepsilon}\right). \quad (9)$$

for the configuration $\vec{E} = E\hat{u}$ and $\vec{B} = B\hat{u}$, where $\hat{u}$ is an arbitrary direction in the real space. Here, $D_{u,n} = D(\vec{B}, \vec{\Omega}_{n,k}) = (1 + e\vec{B} \cdot \vec{\Omega}_{n,k}/\hbar)^{-1}$, $\vec{v}_{n,k} = \nabla_k \varepsilon_{n,k}/\hbar$ is the group velocity and $\varepsilon_{n,k}$ is the quasiparticle energy dispersion, and $\vec{\Omega}_{n,k}$ represents the Berry curvature. $\tau$ is the phenomenological relaxation time and is independent of momentum. $f_{eq}$ is the equilibrium distribution function, and at zero temperature, $(-\partial f_{eq}/\partial \varepsilon) = \delta(\varepsilon - \varepsilon_k)$. For low magnetic fields and away from band touching point $(k=0)$, one can get $D_{u,n} \approx 1$, as shown in Ref. [31]. As a result, $\sigma_{u,0}$ may be anisotropic, but $\gamma_u$ is isotropic.

The longitudinal magneto-resistance can be expressed as

$$MR = \frac{\rho - \rho_0}{\rho_0} = -\frac{\gamma_u B^2}{\sigma_{u,0} + \gamma_u B^2}. \tag{10}$$

We can see that the negative longitudinal magneto-resistance could occur along any arbitrary direction. For low magnetic field, the anisotropy of $MR$ mainly comes from the Drude conductivity $\sigma_{u,0}$.

Above-mentioned discussions reveal that the negative magnetoresistance is extremely anisotropic in the ultra-quantum limit, occurring only along b-axis, and isotropic in the classical limit. One can expect that *in the intermediate quasi-classical region, the negative magnetoresistance should exhibit a crossover from extreme anisotropy to isotropy*. Theoretically, whether a system is in the ultra-quantum limit or semi-classical limit is determined by the product of cyclotron frequency $\omega_c$ and relaxation time $\tau$. The system is in the ultra-quantum limit if $\omega_c \tau \gg 1$, and in the classical limit if $\omega_c \tau \ll 1$ [4, 32]. In our case, the relaxation time $\tau$ is estimated from the resistivity $\rho$ and carrier concentration $n$ ( $\tau = \frac{m}{\rho n e^2}$, $m$ and $e$ are effective mass and charge of carrier, respectively) [32]. The carrier concentration is experimentally determined by two-band fitting of the MR results. The cyclotron frequency is determined by $\omega_c = \frac{eB}{m_e}$ ($B$ is the magnetic field). Combined these equations together, $\omega_c \tau$ is expressed as $\frac{B}{ne\rho}$. *$\omega \tau$, under 5 Tesla magnetic field, is found to be 0.51*. This value is smaller than but comparable to unity. It suggests that our samples are within the quasi-classical region, i.e., *the crossover region from the ultra-quantum limit to the semi-classical limit*. Therefore, it is reasonable to observe in our experiment that non-vanishing but anisotropic negative longitudinal magneto-resistance occur along both the *a* and *b* axes. It should be mentioned that a recent theoretical work [30] also reported the isotropic ABJ anomaly in type-II Weyl semimetal. The main conclusion is quite consistent with our above-discussions.

In summary, we experimentally observe the obviously anisotropic ABJ anomaly in Fermi-level delicately adjusted WTe$_{1.98}$ crystal. Temperature-sensitive ABJ anomaly may be attributed to topological phase transition of WTe$_{1.98}$ under thermal agitation. Theoretical electrical transport study reveals that ansotropic but finite ABJ anomaly can be observed on both *a*- and *b*-axis in type-II Weyl semimetal at the quasi-classical regime.

We would like to acknowledge the financial support from the National Natural Science Foundation of China (51472112, 51032003, 11374140, 11374149, 10974083, 11004094，11134006, 11474150 and 11174127), and major State Basic Research Development Program of China (973 Program) (2015CB921203, 2014CB921103, 2015CB659400). Y.-Y.L acknowledges the financial support from the Graduate Innovation Fund of Nanjing University (2015CL11). S. H. Yao acknowledges the enlightening discuss with Dr. Maxim Avdeev at ANSTO on Rietveld analysis of powder XRD results. The use of the computational resources in the High Performance Computing Center of Nanjing University for this work is also acknowledged.

**Figure Captions**

**FIG. 1. Structural characterizations of WTe$_2$ single crystals.** (**a**) Crystal structure of WTe$_2$: a 3D perspective view along the *a*-axis and a 2D projection along the *c*-axis. W-W chains (highlighted by red-dash line) are formed along the *a*-axis. (**b**) The XRD patterns of representative WTe$_2$ single crystal. Inset is the typical photograph of the as-grown WTe$_2$ single crystals. (**c**) W (red) and Te (green) element mapping images of WTe$_2$ single crystal. (**d**) High resolution TEM image of the WTe$_2$ single crystal. Inset is selected area electron diffraction of WTe$_2$ with electron beam aligned along *c*-axis.

**FIG. 2. Temperature-dependent resistivity along *a*- and *b*-axis of electron-doped WTe$_2$ crystals under varied magnetic field aligned along *c*-axis.** (**a**) The temperature-dependent resistivity $\rho_{xx,\,a}$ with various magnetic fields along the *c*-axis and current along the *a*-axis. Inset is the schematic of the experiment. (**b**) The temperature-dependent resistivity $\rho_{xx,\,b}$ with various magnetic fields along the *c*-axis and current along the *b*-axis. Inset is the schematic of the experiment.

**FIG. 3. Magnetic-field-dependent magnetoresistance of electron-doped WTe$_2$ under different magnetic field directions and different temperatures.** (**a**) The relationship between MR and magnetic field of WTe$_2$ single crystal measured at various temperatures, with $\vec{B} // \vec{j} // a$-axis. (**b**) The relationship between the MR of WTe$_2$ and magnetic field under different θ angles measured at 2 K with current along the *a*-axis. θ changing from 0 to 90° corresponds to magnetic field tilted from *a*- to *b*-axis. Inset shows the negative MR when θ is tilted from 0° to 5°. (**c**) The relationship between MR and magnetic field of WTe$_2$ single crystal measured at various temperatures, with $\vec{B} // \vec{j} // b$-axis. (**d**) The relationship between the MR of WTe$_2$ and magnetic field under different θ angles measured at 2 K with current along the *b*-axis. θ changing from 0 to 90° corresponds to magnetic field tilted from *b*- to *a*-axis. Inset shows negative MR when θ is tilted from 0° to 20°.

**FIG. 4. Analysis of ABJ anomaly in electron-doped WTe$_2$ crystals under different temperatures and magnetic field directions.** (**a**) The magnetoconductance of WTe$_2$

single crystal measured at 2 K under $\vec{B}//\vec{j}//a$-axis or $\vec{B}//\vec{j}//b$-axis conditions. The black or blue circles represent the experimental data and the red dash lines are the fitted results. (**b**) The same theoretical fitting of temperature-dependent ABJ anomaly with $\vec{B}//\vec{j}//b$-axis.

**FIG. 5. Temperature-dependent lattice constants and Weyl-points evolutions at different temperatures in WTe$_2$.** (**a**) Rietveld refinement of the X-ray diffraction data of polycrystalline WTe$_2$ crystal at 35 K. Crosses (×) marks and the red solid lines represent the experimental and Rietveld refinement results, respectively. The differences between the calculated and observed patterns are plotted at the bottom (purple line). The orange vertical lines indicate the calculated positions of the Bragg reflections for the proposed crystal structure. (**b**) Temperature-dependent lattice constants *a*, *b*, and *c*, determined from the experimental X-ray diffraction at 35, 60, 100, 200, and 300 K. The dotted lines are the linear fittings of the experimental lattice constants. (**c**) Band structure of WTe$_2$ with the lattice constants at 0 K. Two crossing points K$_1$ and K$_2$ are the two Weyl points. The positions of the Weyl points can be found at Fig. S3 in supplementary information. (**d**) Same as (**c**) but for the lattice constants at 130 K. A small energy gap appears and no Weyl points are found. (**e**) The whole evolutions of the separation distances between two Weyl points and the energy gaps of WTe$_2$ at different temperatures from 0 to 130 K.

**Table 1** Fitting parameters of negative MR due to ABJ anomaly of electron-doped WTe$_2$ crystal in the case of $\vec{B}//\vec{j}// a$ -axis and $\vec{B}//\vec{j}// b$ -axis conditions.

| | $C_W\ (T^{-2})$ | $a\ (T^{-0.5}(m\Omega\cdot cm)^{-1})$ | $\sigma_0\ ((m\Omega\cdot cm)^{-1})$ | $\rho_0\ (m\Omega\cdot cm)$ |
|---|---|---|---|---|
| $\vec{B}//\vec{j}// a$ -axis | 0.030 | -0.231 | 1.709 | 0.287 |
| $\vec{B}//\vec{j}// b$ -axis | 0.051 | -0.115 | 1.502 | 0.103 |

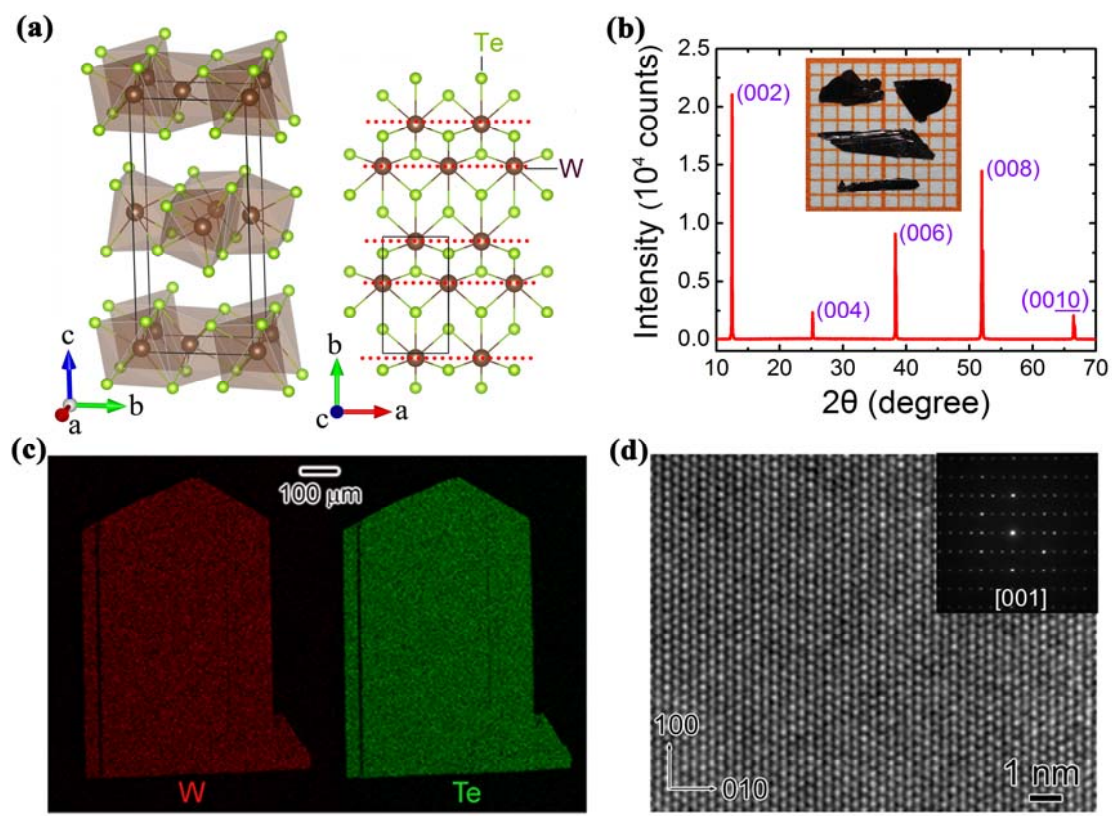

**FIG. 1**

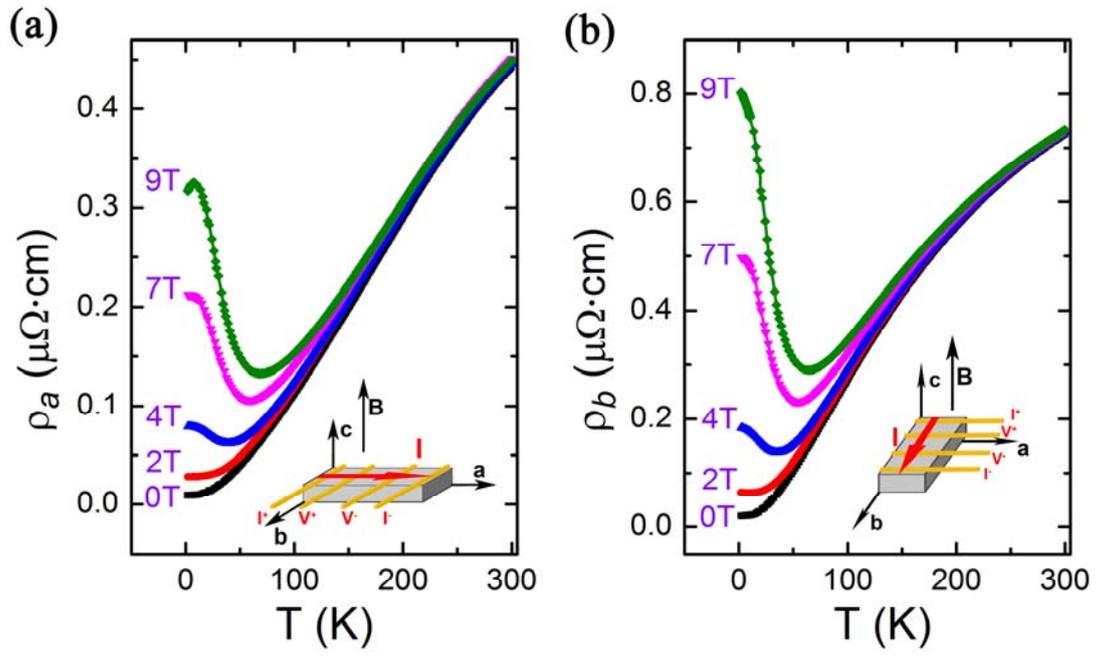

**FIG. 2**

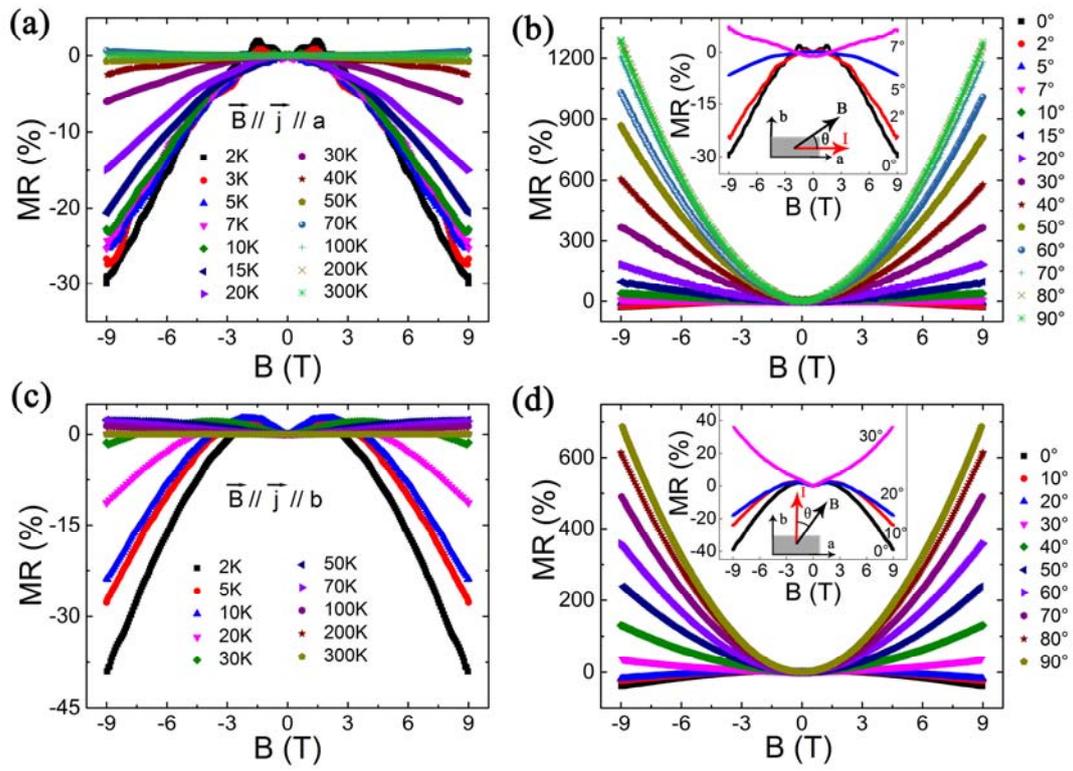

**FIG. 3**

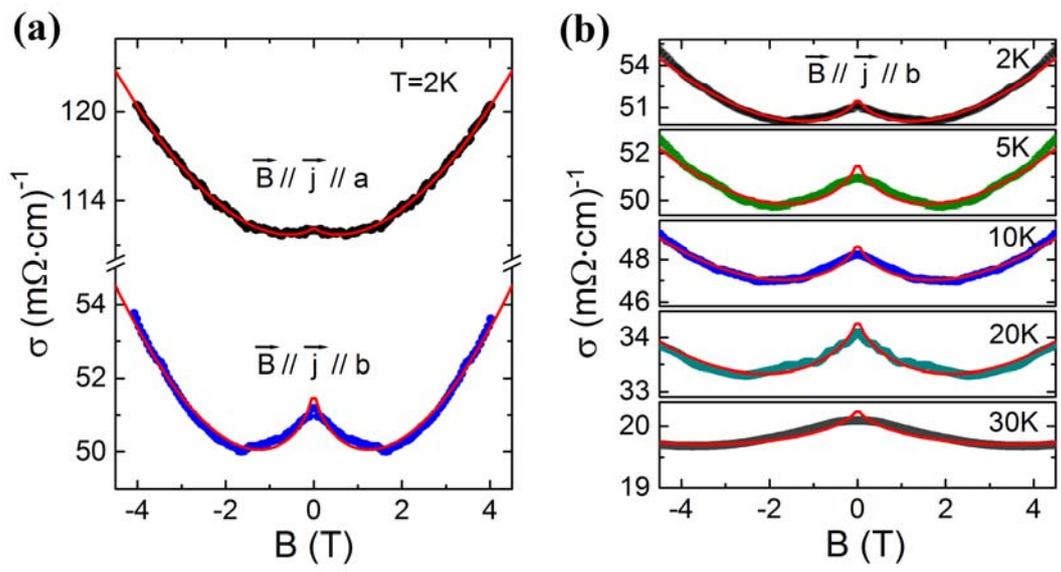

**FIG. 4**

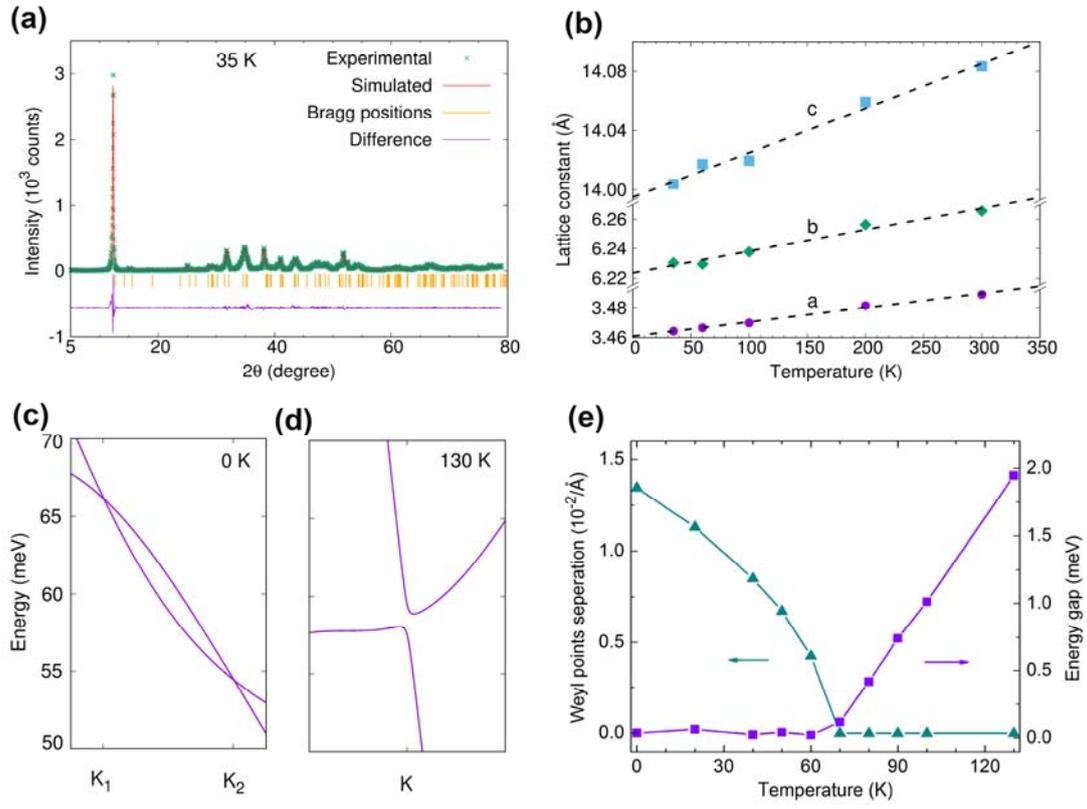

**FIG. 5**